\documentclass[twocolumn]{aastex631}
\usepackage[portuges, english]{babel}
\usepackage[utf8]{inputenc}
\usepackage{amsmath}
\usepackage[makeroom]{cancel}

\usepackage{tikz,hyperref, xcolor}

\definecolor{lime}{HTML}{A6CE39}
\definecolor{Emerald}{HTML}{50c878}
\definecolor{PineGreen}{HTML}{01796F}
\definecolor{ForestGreen}{HTML}{228B22}
\definecolor{Coral}{HTML}{FF7F50}
\definecolor{YellowOrange}{HTML}{E94E16}
\definecolor{sg}{HTML}{0793AC}

\hypersetup{linkcolor=Emerald,citecolor=ForestGreen,filecolor=lime,urlcolor=PineGreen}

\shorttitle{}
\shortauthors{}

\begin{document}

\title{\LARGE Asteroseismology: Looking for axions in the red supergiant star Alpha Ori}

\author[0000-0002-8529-4719]{Clara Severino}
\affiliation{Centro de Astrof\'{\i}sica e Gravita\c c\~ao  - CENTRA, \\
Departamento de F\'{\i}sica, Instituto Superior T\'ecnico - IST,\\
Universidade de Lisboa - UL, Av. Rovisco Pais 1, 1049-001 Lisboa, Portugal}
\email{clara.severino@tecnico.ulisboa.pt}

\author[0000-0002-5011-9195]{Ilídio Lopes}
\affiliation{Centro de Astrof\'{\i}sica e Gravita\c c\~ao  - CENTRA, \\
	Departamento de F\'{\i}sica, Instituto Superior T\'ecnico - IST,\\
	Universidade de Lisboa - UL, Av. Rovisco Pais 1, 1049-001 Lisboa, Portugal}
\email{ilidio.lopes@tecnico.ulisboa.pt}



\begin{abstract}
In this work, for the first time, we use seismic data as well as surface abundances to model the supergiant $\alpha$-Ori, with the goal of setting an upper bound on the axion-photon coupling constant $g_{a\gamma}$. We found that, in general, the stellar models with $g_{a \gamma} \in [0.002;2.0]\times 10^{-10}{\rm GeV}^{-1}$ agree with observational data, but beyond that upper limit, we did not find stellar models compatible with the observational constraints, and current literature. From $g_{a \gamma} = 3.5 \times 10^{-10} {\rm GeV}^{-1}$ on, the algorithm did not find any fitting model. Nevertheless, all axionic models considered, presented a distinct internal profile 
from the reference case, without axions. Moreover, as axion energy losses become more significant, the behaviour of the stellar models becomes more diversified, even with very similar input parameters. Nonetheless, the consecutive increments of $g_{a \gamma}$ still show systematic tendencies, resulting from the axion energy losses. Moreover, we establish three important conclusions:
(1) The increased luminosity and higher neutrino production are measurable effects, possibly associated with axion energy losses. (2) Stellar models with axion energy loss show a quite distinct internal structure. (3) The importance of future asteroseismic missions in observing low-degree non-radial modes in massive stars:internal gravity waves probe the near-core regions, where axion effects are most intense. Thus, more seismic data will allow us to constrain $g_{a\gamma}$ better and prove or dismiss the existence of axion energy loss inside massive stars.
\end{abstract}



\section{Introduction} \label{sec:intro}

Ever since the missing mass problem, pointed out by \citet{Zwicky1933} while examining the Coma cluster, the scientific community has been trying to solve the mystery of the invisible matter that prevents galaxies from flying apart. During the past century, multiple hypothesis for the constitution of dark matter have been proposed and ruled out \citep[e.g.][]{Bertone2018}. Others, such as WIMPS \citep[e.g.][]{Jungman1996}, sterile neutrinos \citep[e.g.][]{Dasgupta2021} and axions \citep[e.g.][]{Duffy2009} are still viable possibilities. This work focuses on the latter. More specifically, it focuses on one of the possible ways that axions can be generated: thermal production in stellar cores of massive stars, which has implications in the structure and evolution of the star.  
To evaluate this effect, we model a red supergiant, using high quality asteroseismic and spectroscopic data.

\subsection{Theory: The PQ mechanism}
\par The axion was originally proposed as a solution to the CP problem of quantum chromodynamics. The lagrangian of strong interactions has a term that allows for violation of symmetry under a charge and parity inversion, which should result in observable effects, such as a neutron electric dipole moment \citep{Abel2020}, scaled by a factor $\Theta \in [-\pi, \pi]$. Nonetheless, even with a sensitivity of the order of $10^{-26}$e cm, this phenomenon has not been detected, constraining the coefficient to $\Theta < 10^{-11}$ \citep{Graham2015}. The strong CP problem sits on the fact that there is no physical reason for $\Theta$ to be negligible. \citet{PecceiQuinn1977} promoted this constant to a dynamical field which is constructed to be the Goldstone component of a U(1) field. The established symmetry spontaneously breaks from which the axion emerges. However, its anomalous character causes the Goldstone boson to couple to the pion, gaining a small potential that drives the axion field towards a minimum, thus preserving CP symmetry. In the chiral limit, the mixing with $\pi^0$ establishes that $m_a f_a \approx m_{\pi} f_{\pi}$ where $f$ is the decay constant \citep{Weinberg1978}. In the case of "invisible" axions, $f_a \gg v_{EW}$, where $v_{EW}$ is the vacuum expectation value of the electroweak scale, making these particles very light and stable \citep{Marsh2016}. Another consequence of the $\pi^0$ mixing is a two-photon vertex \citep{pdg2020}, which makes the axion-photon coupling the main target of axion search experiments where conversion into and from photons is triggered by external electric and magnetic fields, 

\begin{equation}
    \mathcal{L} = g_{a \gamma} \mathbf{E} \cdot \mathbf{B} \phi_a \, ,
    \label{eq:lagrangian}
\end{equation} 

\par where $g_{a\gamma}$ is the axion-photon coupling constant, \textbf{E} and \textbf{B} the electric and magnetic fields, respectively, and $\phi_a$ is the axion field \citep[e.g.][]{pdg2020}.
A broader range of particles relevant to the dark matter problem are Axion Like Particles (ALPs), which similarly to the axion are light pseudo Nambu-Goldstone bosons that emerge from a symmetry break of a U(1) field \citep[see e.g.][]{pdg2020, Yang2017}. These particles, however, do not address the strong CP problem, therefore having no mass constraints set by couplings to the Standard Model particles. Here, we will refer to both axions and ALPs simply by axions. Even though there is no evidence for this class of particles yet, the extensive search for it has set constraints on its properties.

\subsection{Astrophysical Axion Bounds}
\label{sec:AxionBounds}
Axions are light enough to be thermally produced in stellar cores and interact weakly enough to free stream out of them, potentially changing stellar structure and evolution. Hence, several astrophysical experiments have been able to set constraints \citep[see e.g.][]{Graham2015}. By acting as energy sinks, axions accelerate energy production which can further increase neutrino production \citep{Raffelt1990, Raffelt2008}. Measurements of the solar neutrino flux at the SNO collaboration are consistent with  $|g_{a \gamma}| \leq 4.1 \times 10^{-10} \textrm{GeV}^{-1}$ \citep{Vinyoles2015}. Another method is to measure solar axions directly, by converting them back into photons with external magnetic fields. So far, CAST has achieved the best result yet and was able to set the upper limit $ | g_{a \gamma} | < 0.66 \times 10^{-10}$GeV$^{-1}$ for $m_a < 0.02$eV \citep{CAST(2017)}. In the future, more sensitive helioscopes will follow, such as IAXO which is projected to reach the $10^{-12}$GeV$^{-1}$ range \citep{IAXO(2019)}.
Furthermore, it was found that the axion energy losses are most dramatic during the He-burning phase, shortening the time stars spend in this phase of stellar evolution. In this regard, number counts of horizontal branch stars from a sample of 39 globular clusters, when compared with the population of red giants where the Primakoff process is suppressed, were able to set the upper bound $|g_{a\gamma}| < 0.6 \times 10^{-10} \textrm{GeV}^{-1}$ for $m_a \lessapprox 30$keV \citep{Raffelt1999}. Moreover, simulations of stellar models with masses between $8-12M_{\odot}$ showed that the axion cooling can shorten and even suppress the blue loop phase in these stars. Based on these results, the conservative limit on the axion-photon coupling constant was set $|g_{a \gamma}| < 0.8 \times 10^{-10}\textrm{GeV}^{-1}$ \citep{Friedland2013}. A variety of experiments with the purpose of detecting axions and ALPS has been carried out \citep[see e.g.][]{Graham2015, pdg2020}, however, this work focuses solely on the effect axions have in massive stars.%

\subsection{Stellar Structure and Asteroseismology}
\label{sec:Stars}
\par Massive stars are the metal factories of the Universe. In contrast to low mass stars, these involve more extreme physical conditions, enabling fusion into heavier elements. As such, internal mixing processes including convective core overshooting, semiconvection and rotationally induced mixing play a much important role in the evolution of these stars, none of which is fully constrained today \citep{Kippenhahn2012}. Moreover, this variability of internal physical conditions can result in quite distinct paths in the HR diagram \citep{Agrawal2020}. Nonetheless, massive stars are still of interest \citep{Schootlemijer2019, Bowman2020, Cantiello2021}. Here, we are focused on a more evolved stage of stellar evolution, the red supergiant phase, after the H in the nucleus has been extinguished and He fusion is taking place.
\par Bearing in mind that the premise of this work is the impact axions have in massive stars, which is more pronounced in interior layers, spectroscopic measurements alone are not sufficient to draw robust conclusions. Thus, it is necessary to supplement this analysis with seismological measurements and methods. 
Asteroseismology studies the internal structure of stars by analysing their frequency spectra. Depending on the internal conditions of the stars, different modes of oscillation are excited in different regions, thus the measurement and identification of specific modes provides information about its internal physical characteristics \citep{Aerts2010}. That being said, the treatment of stellar oscillations is rather complicated. Its theoretical computation involves solving a system of fourth-order differential equations that result from linear perturbations to the equations of stellar structure \citep{Unno1979}.
Fortunately, some simplifications can be applied without changing significantly the outcome of the original equations, such as the Cowling approximation, that neglects the perturbation to the gravitational potential, resulting in a second order differential equation to describe the radial displacement, $\xi_r$, in function of the angular frequency, $\omega$ \citep{Cowling1941}:  
\vspace{-1mm}
\begin{equation}
    \dfrac{d^2 \xi_r}{d \omega^2} = \dfrac{\omega^2}{c_s^2} \left( 1 - \dfrac{N^2}{\omega^2} \right)\left( \dfrac{S_l^2}{\omega^2} - 1 \right) \xi_r \, ,
     \label{eq:Cowling}
\end{equation}

where $c_s$ is the speed of sound; $S_l$ and $N$ are the Lamb and Brünt-Väisälä frequencies, respectively, which can be defined by the following expressions:
\vspace{-1mm}
\begin{equation}
    S_l^2 = \dfrac{l(l+1)c^2_s}{r^2} \;  {\rm , and }
     \label{eq:lambfreq}
\end{equation}
\vspace{-2mm}
\begin{equation}
    N^2 = g\left( \dfrac{1}{p \gamma_1} \dfrac{d p}{d r} - \dfrac{1}{\rho} \dfrac{d \rho}{d r} \right)^{-1} \, .
     \label{eq:bruntvaisalafreq}
\end{equation}

\par
\begin{figure} 
    \centering
    \includegraphics[width=8.5cm]{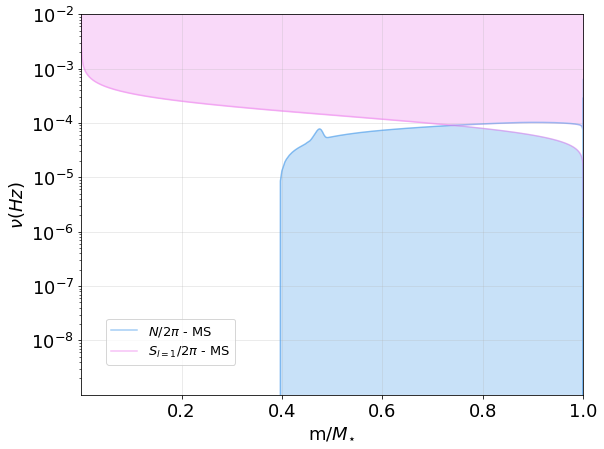}
    \includegraphics[width=8.5cm]{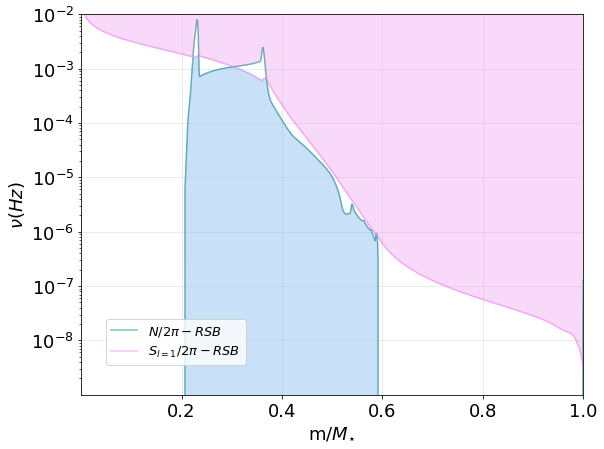}
    \caption{Propagation diagram showing the profile of the Brünt-Väisälä and Lamb (with $l=1$) frequencies for two snapshots in the evolution of Betelgeuse, with mass $M = 16.85M_{\odot}$. The first corresponds to a moment of the MS (top), near ZAMS (age = 193.59 kyr) and the second (bottom) corresponds to a point in the RSB (age = 14.17 Myr). For both, $N$ and $S$ are represented in blue and violet, respectively.}
    \label{fig:NS_example}
\end{figure}

 Where $l$ is the spherical degree of the mode, $p_0$ and $\rho_0$ the local pressure and density in an unperturbed state and $\gamma_1 = \left( \rho \cdot d p /(p \cdot d \rho) \right)_{ad}$ the first adiabatic exponent. Together, these two frequencies govern the oscillatory behaviour of the star (see Fig. \ref{fig:NS_example}). Acoustic modes, where pressure is the restoring force, can only propagate in regions of the star where the frequency, $\nu = \omega/2\pi$, satisfies the condition $\nu >  \left\{ S_l /2\pi, N/2\pi \right\} $ (red and violet regions of Fig. \ref{fig:NS_example}). Analogously, gravity modes, whose restoring force is buoyancy, can only propagate where $\nu < \left\{ S_l /2\pi, N/ 2\pi \right\}$ (orange and teal zones in Fig. \ref{fig:NS_example}). The remaining regions are called evanescent zones because oscillations are damped in these areas. In the illustrated scenarios, one can see how dramatically the interior of a star can change - from an almost fully radiative interior in the Main Sequence (MS) to a pronounced convective core and envelope in the Red Supergiant Branch (RSB).  

\par Therefore, measurements of the stellar periods and posterior identification of the respective modes, provides insight about the region of the star from which the oscillations were generated. Particularly, in the asymptotic regime some simple relations between frequencies emerge. This is the case of the separation in frequency between two consecutive overtone modes and fixed $l$, which is approximately constant and equal to 
\vspace{-1mm}
\begin{equation}
    \nu_{n+1,l} - \nu_{n,l} = \Delta \nu_{n,l} \approx \left( 2 \int_0^{R_{\star}} \dfrac{dr}{c_s} \right)^{-1} \, .
     \label{eq:deltanu}
\end{equation}

\par On the other hand, buoyancy driven pulsations with the same spherical degree, $l$, and consecutive radial order, $n$, are equally spaced in period, with: 
\vspace{-1mm}
\begin{equation}
    \Pi_{n+1,l} - \Pi_{n,l} = \Delta \Pi_{n,l} \approx  \dfrac{2\pi^2}{\sqrt{l(l+1)}}  \left(  \int_{r_1}^{r_2}  N(r)\dfrac{dr}{r} \right)^{-1} \, ,
     \label{eq:deltapi}
\end{equation}

\par where $r_1$ and $r_2$ are the inner and outer boundaries of the g-mode pulsation cavity. These quantities are of fundamental importance, because as basic seismic properties of a given star, they can be measured by means of asteroseismic observational techniques and compared with the equivalent quantities computed for a given stellar model.

\par Over the past decades, space missions with asteroseismic goals have made meaningful contributions to the field, by probing stars continuously for a long time. MOST was the first such mission, paving the way for more complex ones, such as CoRoT, Kepler/K2 and TESS \citep{Mauro2016}. With the exception of the latter, most missions avoid brighter stars due to signal saturation, resulting in more uncertainties in the treatment of massive stars. In the future, PLATO will be launched, expanding further the collection of asteroseismic data \citep{Rauer2021}.

\par In the following sections of this work, we will focus on the modelization of a red supergiant star. The reference model involves only SM physics, while the subsequent ones account for axion energy losses, with different values of $g_{a\gamma}$. Section \ref{sec:modelization} describes the star that will be the object of our study, while the calibration methods are outlined in section \ref{sec:calibration}. The results are then presented in section \ref{sec:results}. Finally, in the conclusion, we summarize the main achievements of this work in addition to future prospects.

\section{Modelization of a red supergiant}
\label{sec:modelization}  

As mentioned earlier, data on massive stars is scarce. With red supergiants, this impediment is scaled by the fact that massive stars evolve faster and their supergiant phase lasts much less time, being more difficult to find. Thus, we focus here on one of the most well known massive stars: $\alpha$-Orionis (M2Iabr), commonly recognized as Betelgeuse.  
Being one of the brightest stars in the sky, Betelgeuse first observations remote back to antiquity when its red colouration was already a staple. Over the centuries, the bright star was noticed multiple times as well as its variability, which, together with limb darkening, different appearances at different wavelengths and an envelope of ejected material around it have made quite difficult to obtain accurate measurements of its radius, distance and parallax \citep{Haubois2009}. Nevertheless, multiple studies on this star have been carried out over the past years \citep[see e.g.][]{Gray2008, Meynet2013, Kravchenko2021}. Some of the most recent results come from \citet{Joyce2020} who arrived at a distance of $168.1_{-14.9}^{+27.5}\,$pc and a radius of $764^{+116}_{-62}\,$R$_{\odot}$ through seismic analysis. The same study used SMEI photometry data to measure the periods of oscillation: $416 \pm 24\,$day ($f_1 = 27.8 \pm 1.6\,$nHz) and $185 \pm 14\, $day ($f_2 = 62.6\pm 4.7\,$nHz) which they respectively identified as the fundamental mode and the first overtone using GYRE \citep{GYRE2013}. Being a semiregular variable of SRc class, $\alpha$-Ori presents an additional long secondary period of $2050 \pm 460\,$day ($f_{LSP}=5.65\pm 1.27\,$nHz), which was determined using light curves collected by the AAVSO \citep{Kiss2006}. Moreover, this supergiant is expected to oscillate in several additional modes excited in its different layers, including gravity waves with a period of $\approx20\,$day, resulting from oscillations of the inner convective regions during core helium burning. However, most waves are likely to be severely dampened before reaching the surface, specially due to shock and thermal dissipation \citep{Nance2018}.

\par One of the main focuses of Betelgeuse studies is to understand in which phase of stellar evolution this star is and how long it will be until a supernova occurs. \citet{Lambert1984} concluded that Betelgeuse was past the first dredge-up phase, based on the surface abundance of Carbon, Nitrogen and Oxygen which were consistent with the CN-processed material being mixed to the surface. Moreover, most models built point to a present day star ascending the RSB, while undergoing He core fusion \citep{Dolan2016, Joyce2020, Neilson2011}. Nevertheless, most cases do not account for the exceptionally fast rotation of $\alpha$-Ori. An exception is \citet{Wheeler2017}, however, their rotating models do not give reasonable evolutionary predictions. Most converge to much lower rotation velocities, while the few that are able to achieve such values, are only able to do so for a brief moment in the bottom of the RSB.  
As a consequence, a past merger event was suggested as the cause for the high rotation velocity \citep{Chatzopoulos2020, Sullivan2020}. If this is the case, initial conditions for Betelgeuse become much more complex to calibrate. One such case is the initial mass, usually assumed to be between $15M_{\odot}$ and $25M_{\odot}$. Thus, the present day mass provides a more relevant aspect, which \citet{Joyce2020} placed in the range $16.5-19M_{\odot}$ based on a seismic analysis.     

Here, we make use of the available spectroscopic and seismic data to model $\alpha$-Ori while testing the hypothesis of axion production in its core. It is not the first time Betelgeuse has been the object of dark matter searches; recently \citet{Xiao2021} examined its hard X-ray spectrum, measured by the NuSTAR satellite telescope, to evaluate the production of ALPs which would subsequently convert into photons. As a result, they were able to set the upper limit $|g_{a\gamma}| < (0.05-0.18) \times 10^{-10}\textrm{GeV}^{-1}$ for $m_a < (5.5-3.5) \times 10^{-11}$eV (depending on the magnetic field model).

\section{Calibration}
\label{sec:calibration}

\par Observations aside, stellar modelization is the best tool in the study of stars and the physics involved in their interiors. In this field, the Modules for Experiments in Stellar Astrophysics (MESA) \citep{Paxton2011, Paxton2013, Paxton2015, Paxton2018, Paxton2019}, an open-source, one dimensional stellar evolution module, stands out by its robustness, efficiency and applicability, as it solves the fully coupled structure and composition equations for a wide range of scenarios. One such case is the introduction of axions, developed by \citet{Friedland2013} and corrected by \citet{Choplin2017}, which we used in this work. The axion energy losses were computed as 
\vspace{-1mm}
\begin{equation}
    \varepsilon_{ax} = 283.16 \times g_{10}^2 T_8^7 \rho_3^{-1} \xi^2 f(\xi^2) \quad \textrm{erg/g/s} \, ,
     \label{eq:axionenergyloss}
\end{equation}

where $g_{10} =  10^{10} g_{a\gamma}\,$GeV, $T_8 = 10^{-8} T\, $K$^{-1}$, $\rho_3 = 10^{-3} \rho \, $(g/cm$^3$)$^{-1}$, $f$ is defined in eq. (4.79) of \citet{Raffelt1990} as an integral over the photon distribution, and $\xi$ is given by
\vspace{-1mm}
\begin{equation}
    \xi = \dfrac{\hbar \, c \, \kappa_s}{2 k_B \, T} \, ,
    \label{eq:xi}
\end{equation} 

where $\hbar$ is the reduced Planck constant, $c$ the speed of light, $k_B$ the Boltzmann constant and $\kappa_s$ the Debye-Huckel screening wave number, presented in \citet{Raffelt2008}: 
\vspace{-1mm}
\begin{equation}
    \kappa_s^2 = 4 \pi \alpha \left( \dfrac{\hbar c}{k_B T} \right) \sum_{\textrm{i=e,ions}} n_i Z_i^2 \, ,
     \label{eq:kappa2}
\end{equation}

with $\alpha$ being the fine structure constant, $Z_i$ the atomic number and $n_i$ the number density: 
\vspace{-1mm}
\begin{equation}
    n_{\textrm{ions}} = \rho \dfrac{X_i}{A_i} \mathcal{N}_A \quad \textrm{and} \quad n_e = \sum_{\textrm{i=ions}} n_i Z_i \, ,
    \label{eq:numberdensity}
\end{equation} 

where $\mathcal{N}_A$ stands for the Avogadro number, $A_i$ the molar mass and $X_i$ the weight fraction of each nuclei. In this work, we applied these calculations to stellar models with $g_{10} \in [0.002;\, 5.0]$ in different increments.

\par In addition to the axion energy loss calculations, the calibration of Betelgeuse made use of the \textit{astero} module \citep{Paxton2013}, responsible for implementing algorithms that search for model parameters that best match observations. The module accepts a few selected input parameters: $\{M, Y_i, [Fe/H]_i, \alpha, f_{ov} \} \equiv \,$\{mass, initial helium abundance, initial metallicity, mixing length parameter, overshooting parameter\}, based on which it evolves stellar models that try to minimize the $\chi^2 $ to the observational data. Here, we used the quantities presented in Table \ref{tab:input} for this purpose. The effective temperature, $T_{eff}$, was taken from \citet{Levesque2020}, the surface abundances, $\{X(^{12}C),\, X(^{14}N),\, X(^{16}O)\}$, from \citet{Lambert1984} and the remaining spectroscopic, $\{\log(L/L_{\odot}),\, R/R_{\odot}\}$, and seismic quantities, $\{ f_1 ,\,f_2\}$, from \citet{Joyce2020}. Regarding the diagnostic parameter $\chi^2$, it was calculated using the standard weighted contributions of $\chi^2_{\textrm{seismo}}$ and $\chi^2_{\textrm{spectro}}$ \citep{Metcalfe2012, Paxton2013}: 
\vspace{-1mm}
\begin{equation}
    \chi^2 = \dfrac{1}{3} \chi^2_{\textrm{spectro}} + \dfrac{2}{3} \chi^2_{\textrm{seismo}} \, ,
    \label{eq:chi2weight}
\end{equation} 

which quantify the deviations to the seismic and spectroscopic observed data, respectively. Each one is calculated as
\vspace{-1mm}
\begin{equation}
    \chi^2_{\textrm{seismo/spectro}} = \dfrac{1}{N} \sum_i \left( \dfrac{y_i^{\textrm{obs}} - y_i^{\textrm{model}} }{\sigma_i} \right)^2 \, ,
     \label{eq:chi2}
\end{equation}

where $N$ is the number of parameters, $y^{\textrm{obs}}$ the observed value, $y^{\textrm{model}}$ the corresponding match of each model and $\sigma$ the observational uncertainty. The $\chi^2_{\textrm{seismo}}$ incorporated $f_1$ and $f_2$, whereas the $\chi^2_{\textrm{spectro}}$ included the usual quantities; the luminosity, effective temperature and radius, as well as, the more uncommon, surface abundances of carbon, nitrogen and oxygen.

 \begin{deluxetable*}{c c c c c c c c }
     \centering
    \label{tab:input}
      \tablecaption{Input parameters used in $\alpha$-Ori calibrations, based on observations.}
     \tablehead{\colhead{\textbf{Log(L$/L_{\odot}$)}} & \colhead{$\mathbf{R/R_{\odot}}$}  &  \colhead{$\mathbf{T_{eff}}$ (K)} & \colhead{$\mathbf{f_1}$(nHz)}  & \colhead{$\mathbf{f_2}$(nHz)} & \colhead{$X(^{12}C)$ ($10^{-3}$)} & \colhead{$X(^{14}N)$ ($10^{-3}$)}  & \colhead{$X(^{16}O)$ ($10^{-3}$)} }
     \startdata
        $4.94_{-0.06}^{+0.1}$ & $764_{-62}^{+116}$ & $3600\pm25$ & $27.8\pm 1.6$ & $62.6\pm4.7$  & $1.85\pm0.80$ & $4.1\pm 1.1$ & $6.6\pm2.0$  \\ 
    \enddata
 \end{deluxetable*}

\par In terms of the calibration itself, the downhill simplex algorithm was implemented to minimize the $\chi^2$ \citep{Nelder1965}, which has proven to be a reliable method, given its ability to adapt to the current landscape towards finding an absolute minimum. 
\par In addition, all stellar models were developed from pre-main sequence and neglected the effects of rotation and magnetic fields. Moreover, we employed the ‘Dutch’ mass-loss prescription with $\eta = 0.8$, which is consistent with the $\Dot{M} = (2 \pm 1) \times 10^{-6} M_{\odot}/{\rm yr}$ adopted by \citet{Dolan2016}. In order to investigate the impact of this parameter in the evolution of massive stars, we generated uncalibrated models with values of $\eta$ between $0.6$ and $2.4$. Among all cases, we did not find a significant impact on the stellar structure, and thus kept this parameter fixed for all calibrated models. Lastly, we employed an exponential overshooting scheme, similarly to \citet{Wheeler2017} and \citet{Joyce2020}. 

\section{\texorpdfstring{$\chi^2$}{TEXT} analysis}
\label{sec:chi2}

\par Before we analyse the effects of axion energy losses in a red supergiant, we start by noting that not only was the simplex algorithm successful in the search for calibrated models with $g_{10} < 3.5$, it was also able to return hundreds of plausible models within this range. Given the large uncertainties of the observational data, there is enough leeway to generate numerous fitting models. However, since the calculated $\chi^2$ has contributions from each individual observational parameter, by analysing the models more carefully, we have noticed that for a significant percentage of them, the $\chi^2$ is dominated by one particular variable. Thus, an additional criterion was implemented: that each term in the sum of eq. (\ref{eq:chi2}) should be smaller than 2. With the ensemble of models reduced, but still significant, the possibility of over fitting the data needs to be addressed. To guarantee that the results presented in the following sections are not arbitrary, a statistical analysis was conducted to make sure that in each line of Table \ref{tab:modeloutputs} there is a stellar model representative of that class of axion energy losses. In this respect, a random sample of models for each $g_{10}$ was selected. To assess the compatibility of each sample of models, the stellar profile at the moment of convergence was compared in ten categories: \{nuclear energy production, temperature, density, Lamb and Brunt-Väisälä frequencies, and the mass fraction abundances of $^1H$, $^4He$, $^{12}C$, $^{14}N$ and $^{16}O$\}. In addition, the evolution of the luminosity resulting from triple-$\alpha$ reactions was evaluated as well, whose plots matched almost perfectly for most stellar models in each sample; while displaying a distinct path when $g_{10}$ is varied, as will be discussed in section \ref{sec:results:AEL}.
\par The analysis of the converging profiles is somewhat more complicated, given that a slight variation of the initial parameters can have drastic effects, including the moment at which the model converges, changing the corresponding profile. An important remark to make is that for $g_{10} \in [0.002; 0.06]$, models undergoing He-fusion in a shell around the core were found, but will not be considered for this analysis. Such phenomenon was only observed for models with $X(^{14}N)\geq 3.85 \times 10^{-3}$, which was not obtained for the remaining cases, but it is possible that the $\chi^2$ minimisation process incurred in a local minimum and such models do exist.   

\begin{figure}
    \centering
    \includegraphics[width= 9cm]{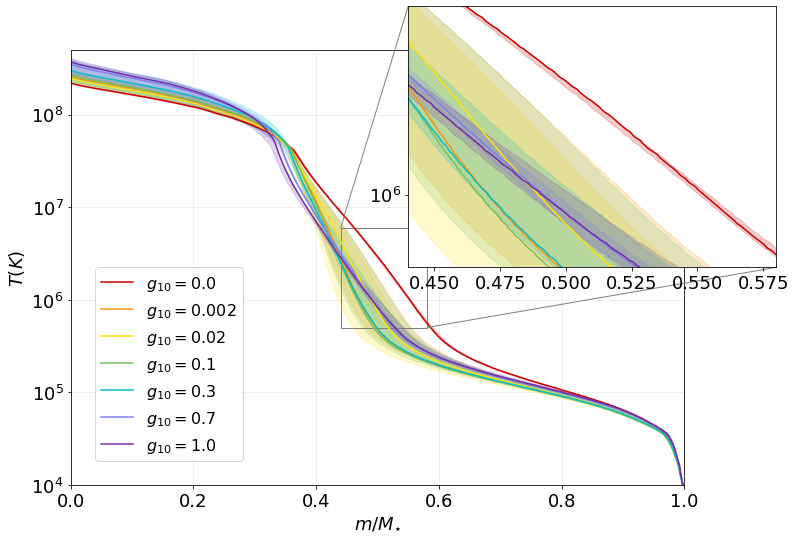}
    \caption{ Temperature profile of the $\alpha$-Ori models. The solid lines correspond to the model that best represents the average behaviour for the corresponding $g_{10}$, while the shaded regions illustrate the uncertainty. } 
    \label{fig:temperature}
\end{figure}

\par Nonetheless, the analysis of the randomly selected samples shows that the cases without axions are the most consistent and diverge, in all aspects, from the models with axions. An example is shown in Fig. \ref{fig:temperature}, for the temperature profile. As we introduce axion energy losses, the lower values of $g_{a \gamma}$ give rise to similar models, but as the axion energy loss channel becomes more significant, so does the variability between models, thus increasing the possibility of degenerate models, specially if one considers stellar models with $g_{10}$ very closely spaced. Nonetheless, for the increments considered in this work, the overlap of the uncertainty regions (as illustrated in Fig. \ref{fig:temperature}), does not occur among all profile categories examined, and a clear increase in the influence of the axion energy losses is evident as the coupling parameter increases.


\section{Results}
\label{sec:results}

\par Having selected the stellar models that best represent the behaviour of the star, for each $g_{a\gamma}$, the corresponding relevant data can be found in Tab. \ref{tab:modeloutputs}.
In particular, one can notice the range of the input parameters along each column, meaning not only that the observational data is not restrict enough to produce a specific type of model, but also that the introduction of axions might have an impact that reaches all these different characteristics. The respective uncertainties were evaluated by combining the calculation of the standard deviation of the sample of models that obey the two conditions described in the previous section, with the $95\%$ confidence ellipses that correlate these parameters \citep{Delholm2012}. We noted that theses values remain more or less constant despite the corresponding $g_{10}$ parameterization -- hence the inclusion of one value for each column, for which we adopted the highest value. Moreover, to check the  consistency of our results with the current literature, we have performed an additional  axionless calibration of $\alpha$-Ori, without the surface abundances as observational constraints -- given that these quantities are not considered in most studies -- from which we obtained the input parameters of: $M_i = 15.6 \pm 3.1  M_{\odot}$, $Y_i = 0.28 \pm 0.07$, $[F_e /H]_i = -(6 \pm 3) \times 10^{-3}$, $\alpha = 1.9 \pm 0.2$ and $f_{ov} = 0.029 \pm 0.01$. In general, these results are consistent with the values adopted/estimated in other studies \citep{Joyce2020, Dolan2016, Neilson2011}. 
 \par At this point, it is also important to note that the main goal of this calibration was to generate models that best describe the present-day $\alpha$-Ori. With this purpose, the input parameters were given enough freedom so that they could eventually converge to unlikely values. For instance, the initial He abundance is lower than the cosmological He abundance \citep{Planck2016}. We consider this to be a strength of this work, given the uncertainties associated with the evolution of massive stars. Indeed, if different specifications are adopted in the evolutionary code, it is likely that these values change, while preserving the output data. Nevertheless, given the specifications on the physics adopted for this work, the small window of variability in the input parameters, is a consequence of the requirement that the present-day stellar model has surface abundances close to the observed values. For reference, rotation \citep{Luo2022}, convection and mass loss, are important factors that affect the theoretical surface abundances -- among other characteristics -- thus controlling the flexibility to match the target values.

\begin{deluxetable*}{c|ccccc|ccccccccccc}
    \tablecaption{$\alpha$-Ori models. Each one is identified by its axion-photon coupling paramter $g_{10}$. In the following columns there are the input parameters, the age of the model, as well as the best matches for the observed parameters, the $\Delta \Pi$, and finally the $\chi^2$. }
    \tablewidth{\textwidth}
    \tabletypesize{\scriptsize}
    \centering
    \label{tab:modeloutputs}
    \tablehead{ \colhead{\textbf{Model}} & \colhead{\textbf{M}} & \colhead{Y$_i$} & \colhead{$\mathbf{[Fe/H]_i}$} & \colhead{$\mathbf{\alpha}$} & \colhead{\textbf{f}$_{ov}$} & \colhead{\textbf{log}($L/L_{\odot}$)} & \colhead{\textbf{R}} & \colhead{\textbf{T}$_{eff}$} & \colhead{\textbf{Age}} & \colhead{\textbf{f}$_1$} & \colhead{\textbf{f}$_2$} & \colhead{$\mathbf{X(^{12}C)}$} & \colhead{$\mathbf{X(^{14}N)}$} & \colhead{$\mathbf{X(^{16}O)}$} & \colhead{$\mathbf{\Delta \Pi}$} & \colhead{$\mathbf{\chi^2}$} \\
    \colhead{g$_{10}$} & \colhead{$(M_{\odot})$} & \colhead{} & \colhead{$(10^{-3})$} & \colhead{} & \colhead{$(10^{-2})$} & \colhead{} & \colhead{$(R_{\odot})$} & \colhead{(K)} & \colhead{(Myr)} & \colhead{(nHz)} & \colhead{(nHz)} & \colhead{$(10^{-3})$} & \colhead{$(10^{-3})$} & \colhead{$(10^{-3})$} & \colhead{(min)} & \colhead{} 
} 
    \startdata
    0.0  & 16.85 & 0.222 & -7.091  & 1.952  &  2.850  &  4.9053  & 729.0  &  3602.6  & 14.168  & 27.77 & 61.84  &  2.162 &  2.716 &  6.971 &  24.81 & 0.1284 \\
     0.002  &  17.08  & 0.219  &-7.304  & 2.016  &  2.432  &  4.9074  &  728.7   &  3607.5  &  13.960  &  28.00   & 60.97 &  1.884  &  3.404  &  6.602  &  20.04  &  0.0913  \\ 
    0.006  &   17.08  &  0.221  & -7.111   &   1.996   &  2.193   &  4.9040  & 730.8  &  3595.5  &  13.634  &  28.09  &  60.98  &  1.972  &  3.054  &  6.827  &  20.93  &  0.1174 \\
     0.01  &  17.49  & 0.217  &  -6.251   &  1.978 & 2.276  &  4.9178   &  738.2  &  3605.6  &  13.549 & 27.94  & 61.04  & 1.983  & 3.266  &  6.693  &  21.63  &  0.0843  \\
     0.02  & 16.69  & 0.219  &  -6.661   & 2.016  & 2.418  & 4.8982    & 725.6   &  3596.1  & 14.380  &  27.98 & 61.17  & 2.043  & 2.948  &  6.933 &  20.91  &  0.1319  \\ 
     0.06  &  17.08 &  0.219  &  -6.681   & 2.016  & 2.425  &   4.9086  & 729.7    &  3607.6  &  13.913  & 28.03  & 60.83  &  1.839  & 3.541  & 6.524   &  18.78  &  0.0908 \\
     0.1  & 16.24  & 0.239 &  -6.891  & 1.979 & 2.535  & 4.8905   & 718.8   & 3597.3 & 13.940 & 27.91 &  61.05 & 1.904 & 3.151  & 6.544  & 17.87 &  0.1218 \\
     0.2  & 16.78  & 0.227  &    -7.548  &  1.959 & 2.908  &  4.9128   & 735.9   &  3600.9  & 13.920 & 27.66  &  61.26 & 2.183 & 2.572  &  7.020 &  19.52  & 0.1614  \\
     0.3  & 16.73  & 0.240  & -5.662    & 1.972  & 2.909  &  4.9214   &  743.6  & 3600.1   & 13.273 &  27.16  & 59.59 & 1.951  & 3.086  &  6.580  & 13.13 & 0.2446  \\ 
     0.4  &  16.60 & 0.238  &   -7.565  &  1.977 &  2.913 &  4.9020   &  714.0  &  3633.2  &  13.376  &  29.91 & 64.47  & 2.113 & 2.651  &  6.830 &   20.26 & 0.8589  \\
     0.5  & 16.90  & 0.249  & -7.294 &  1.923 & 3.097  &  4.9235   &  740.8  & 3611.3   & 12.608 &  28.03 & 61.30  & 2.016 &  2.822 &  6.598 &  16.14  & 0.1261  \\ 
     0.6  &  16.71 & 0.251  &  -7.513   &  1.878 &  3.309 & 4.9199    & 747.2   & 3588.1   & 12.712  & 27.42  & 60.39  & 2.032 &  2.799 &  6.548 & 17.08   & 0.1922  \\
     0.7  & 16.47  & 0.261  &  -5.537   & 1.897  &  3.399 &  4.9198   & 739.2   &  3607.3  & 12.538 &  27.92 &  61.27 & 2.051 & 2.676  & 6.562  &    9.43 &  0.1374 \\
     0.8  & 16.57  &  0.245 &   - 6.661 & 1.903  & 3.379  & 4.9180 & 743.2   &  3594.0  & 13.108 & 28.12  & 61.33  & 2.108 & 2.685  & 6.717  &  15.17  & 0.1471  \\
     1.0  & 16.84  & 0.256  &   -7.086  &  1.837 & 3.584  &  4.9304   &  756.3  & 3588.3   & 12.348 & 27.42  & 60.47  & 2.081 & 2.691  &  6.544 &  9.00  & 0.1965  \\
     2.0  & 16.78  &  0.253 &  -7.431   &  1.847 & 3.799  &   4.9367  &  763.4  &  3584.5  & 12.543 & 27.43  & 60.23  & 2.137 & 2.578  & 6.646   &  17.61  & 0.2373  \\ \hline
      &  $\pm$0.1 & $\pm$0.004 &  $\pm$0.2 & $\pm$0.02 & $\pm$0.08 &  & & & & & & & & & \\
    \enddata
      \end{deluxetable*}
\subsection{Axion energy losses} \label{sec:results:AEL}

\par
\begin{figure}[ht!]
    \centering
    \includegraphics[width = 8.5cm]{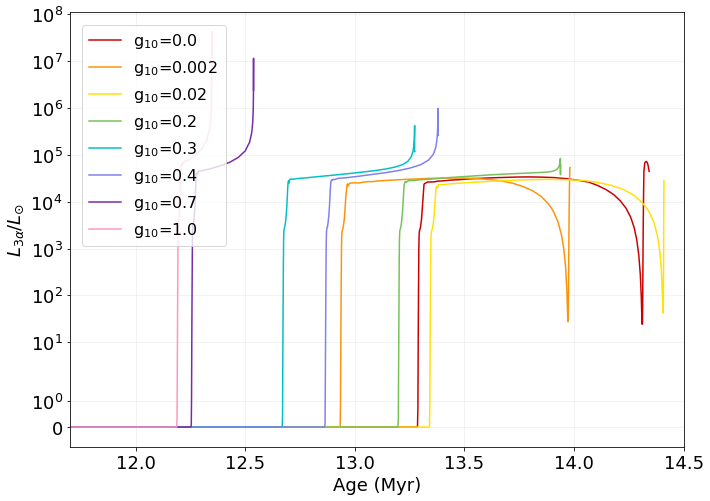}
    \caption{Luminosity of the $\alpha$-Ori models that result from the triple-$\alpha$ reactions. }
    \label{fig:3alpha}
\end{figure}

 Nevertheless, one effect already examined in previous work \citep{Friedland2013, Raffelt1990, Raffelt2008} is the acceleration of He-fusion, that is triggered to compensate the additional energy loss channel. In Fig. \ref{fig:3alpha}, the luminosity resulting from the triple-alpha reactions is represented for a few selected models. The first thing to point out is that all the evolved models are undergoing He-fusion, in accordance with the current literature. A more interesting point is how the luminosity evolution changes with increased axion energy losses. For models with $g_{10}< 0.2$ the minimum after the plateau is attenuated when compared with the axionless case. Moreover, for the complementary models ($g_{10}\geq 0.2$) the minimum completely disappears and the luminosity starts to increase exponentially instead. For $g_{10}\geq0.7$ even the plateau stage is suppressed.
Given that this effect is observed in every model in a systematic manner, independently of the input parameters, it must be a direct consequence of the axions. 

\par
\begin{figure}[ht!]
    \centering
    \includegraphics[width = 8.5cm]{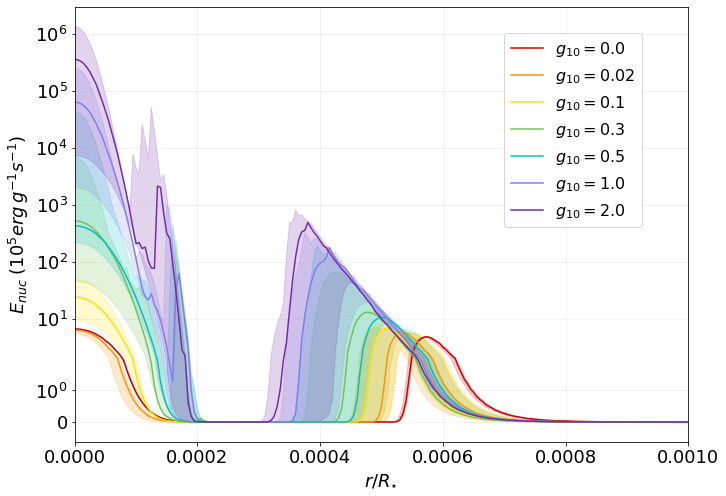}
    \caption{ Nuclear energy produced in the core and near-core regions of the $\alpha$-Ori models. Similarly to Fig. \ref{fig:temperature}, the shaded regions represent the uncertainty associated with the models. }
    \label{fig:Enuc}
\end{figure}

 On the other hand, Fig. \ref{fig:Enuc} shows the nuclear energy produced in the most inner region of the star, which does not present the same behaviour. The energy loss due to neutrinos mirrors the behaviour of the nuclear energy produced, nonetheless, it is not sufficient to create the pattern observed in Fig. \ref{fig:3alpha}. Therefore, we infer that the mechanisms associated with the energy transport between the He-nucleus and the surface might be affected by the axion's existence as well. Moreover, the uncertainty regions illustrate how diverse the profiles can be at the moment of convergence, particularly in higher axion production models. Given that this point occurs in a rapid changing phase of stellar evolution, which is fastened by the presence of axions, even the slightest variations in the input parameters cause dramatic changes.

\par In Fig. \ref{fig:MF+NS}, one can see how the internal structure of the star changes, however, the impact of the axion energy losses is so dramatic that several processes compete with each other leading to a pretty complex pattern of evolution, different for each combination of parameters. 

\par Despite the aforementioned heterogeneity, we divide our models into three categories based on the achieved results: in the first, we consider a low impact of axions with $g_{10} \leq 0.06$; the second one involves a high impact of the axion energy losses, with $g_{10} \geq 0.5$. The intermediate cases are the ones that do not present the characteristics of the other categories, with $g_{10} \in \, ]0.06, 0.5[$. In most cases, the stellar models are able to converge to a point closer to the observational data at a younger age than the control model, probably due to the faster rate at which He is being consumed, that for $g_{10}\geq 0.2$ even competes with the extended MS lifetime prompted by the increased convective core overshooting. This aspect is characteristic of the high impact regime to a point that it contributes to the higher luminosity along the evolution of the star. Moreover, overshooting tends to minimize the effects of the first dredge-up, thus enhancing the discrepancies of the surface abundances with observations \citep{Dolan2016}. In this regime, the effects of the axions are so intense that they manifest throughout the whole star, affecting the luminosity and radius as well. 

\par On the other hand, when we consider the low axion impact regime, the parameters that change the most are the surface abundances of the CNO elements, in particular the nitrogen surface abundance, and the overshooting parameter, which in contrast, decreases. In the top right panel of Fig. \ref{fig:MF+NS}, one can see that the He fusion is in a more advanced stage since the most inner region is already depleted of He, supporting the idea that axions accelerate stellar evolution, specially after the main sequence.

\subsection{Seismic analysis}

\par
\begin{figure*}[ht!]
    \centering
    \includegraphics[width = 8.5cm]{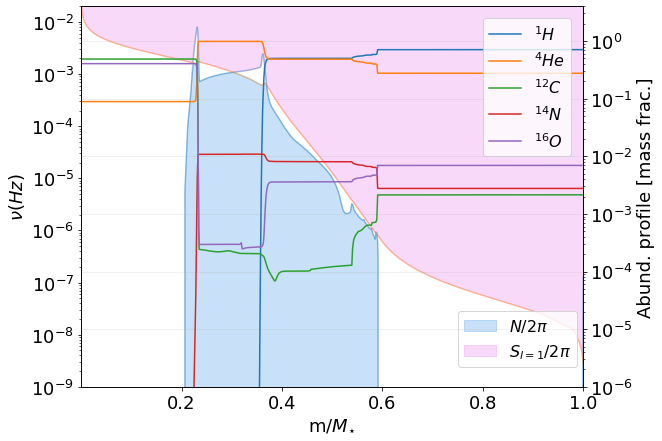}
    \includegraphics[width = 8.5cm]{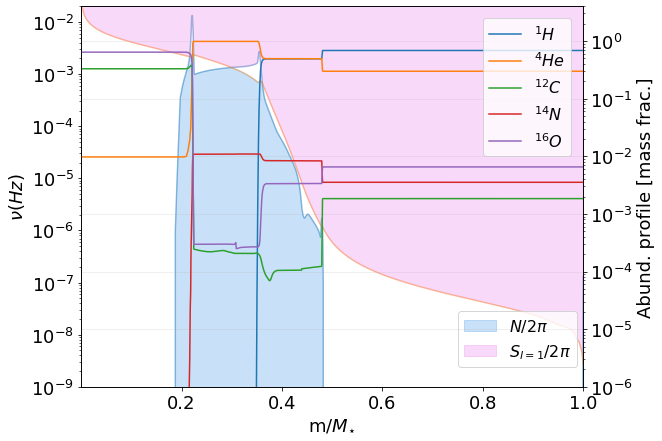}
    \includegraphics[width = 8.5cm]{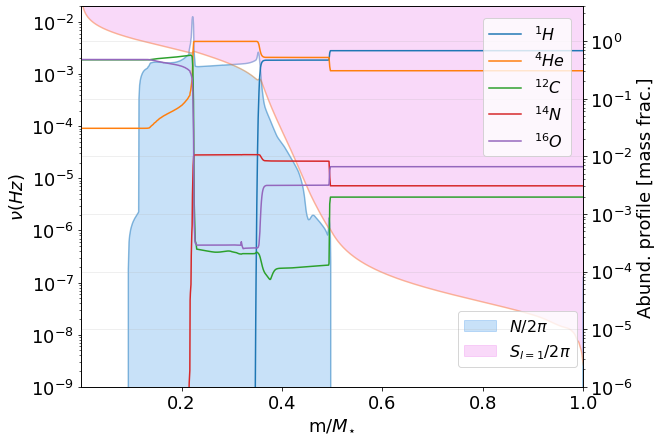}
    \includegraphics[width = 8.5cm]{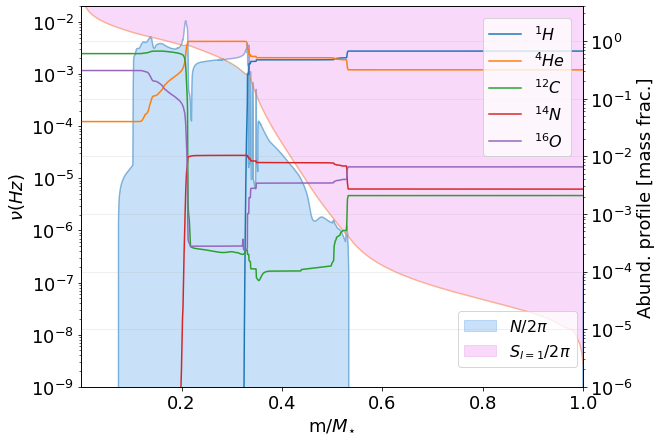}
    \caption{Lamb and Brunt-Väisälä frequencies (left axis), and mass fraction abundances (right axis) for the $\alpha$-Ori model with $g_{a \gamma} = 0.0\, \textrm{GeV}^{-1}$ -- reference model -- (top left panel); $g_{a \gamma} = 0.06 \times 10^{-10}\, \textrm{GeV}^{-1}$ (top right panel); $g_{a \gamma} = 0.3 \times 10^{-10}\, \textrm{GeV}^{-1}$ (bottom left panel) and $g_{a \gamma} = 1.0\times 10^{-10}\, \textrm{GeV}^{-1}$ (bottom right panel).}
    \label{fig:MF+NS}
\end{figure*}

 Still in Fig. \ref{fig:MF+NS}, it is evident that the profile of the Lamb and Brünt-Väisälä frequencies changes with the introduction of axions. In all cases considered, the radiative region ($N^2 > 0$), between the two convective ones, is suppressed on the outer side, which has effects on other aspects of the star, such as the temperature profile (see Fig. \ref{fig:temperature}).  
\par
\begin{figure}[h!]
    \centering
    \includegraphics[width = 8.5cm]{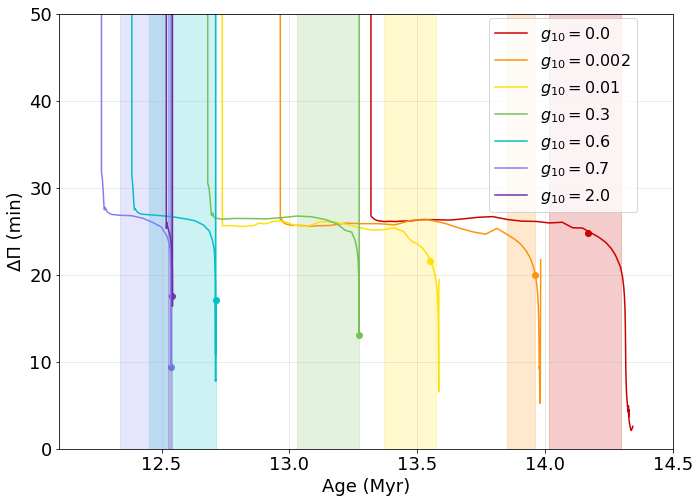}
    \caption{ Evolution of the period spacing $\Delta \Pi$ as a function of time. The corresponding coloured points, mark the moment of convergence.}  
    \label{fig:deltapi}
\end{figure}

 Compared to the reference model, the high axion impact category comprises stellar models with an enlarged and more irregular radiative region, indicating that a more pronounced axion presence inhibits convection. In addition, the He core is partially radiative, with a second incidence of He fusion in these conditions (see Fig. \ref{fig:Enuc}). Furthermore, gravity driven pulsation modes probe the regions near the core and are sensitive to the presence of chemical gradients which cause spikes in the Brunt-Väisälä frequency, leading to mode trapping \citep{Pedersen2018}. Thus, a possibly good indicator of this phenomenon is the $\Delta \Pi$. In Fig. \ref{fig:deltapi}, the evolution of this quantity through time can be found, as well as the points towards which the evolved models converged to the observational parameters. At first glance, one can notice how this evolution mirrors Fig. \ref{fig:3alpha}, especially in the fact that an abrupt variation is followed by a near constant phase, that is shortened by the presence of axions -- in a systematic manner. Moreover, we identify three phases where the convergence can occur: the first is the plateau phase, where $\Delta \Pi \approx 26\, $min (top left panel of Fig. \ref{fig:MF+NS}); the second is along a descending path, corresponding to the retraction of the convective core (bottom left panel of Fig. \ref{fig:MF+NS}); and the third is a sudden rise that can reach past $1000\,$min. This last phase is an indication of a second dredge-up, that has been linked to the second incidence of He-fusion, and in the higher $g_{10}$ cases, to the onset of C-fusion. Nonetheless, the amount of fitting models obtained for each $g_{10}$ parameterization, made possible by the large uncertainties of the observational parameters, does not enable us to set strict limits on which phase the models converge to. Therefore, there is still a lot of work to be done to constrain Betelgeuse and massive stars in general, however, if more modes are discovered -- specially mixed modes --, even with small amplitudes, it will be possible to derive a $\Delta \Pi$ value from observations. Even so, the first two phases are within the range of the third. Hence, the detection of internal gravity waves with the characteristic periods of He-core, He-shell and C-core fusion mentioned in \citet{Nance2018}, will allow us to better constrain the evolutionary state of Beteleguese. Moreover, the detection of these waves, together with the already known CNO element abundances will be able to constrain the mixing mechanisms involved in this star \citep{Pedersen2018}. Such achievement is expected in the future with the PLATO mission, given that it will also probe stars of this brightness. However, this will only be possible if lower magnitude and high frequency modes, as the ones mentioned in \citet{Nance2018}, are detected, considering that most are expected to fall into the asymptotic regime.  For now, as a proof of concept, we perform a calibration of Betelgeuse with $\Delta \Pi = 15 \pm 1.5\,$min as an additional observational parameter.  Note that, these values were chosen to belong to the descending path of $\Delta \Pi$, to show the diagnostic potential of such measurement. Moreover, the uncertainty of $10\%$ was chosen as a conservative value, much higher than what is possible for low mass stars. For the chosen values, the calibration of a reference model ($g_{10}=0.0$) was not able to find models that satisfy the conditions in Section \ref{sec:chi2}. On the other hand, models in the high axion impact had no trouble reaching the target $\Delta \Pi$ range, but once again struggled to reach the $X(^{14}N)$, given that the mixing in the envelope is not able to accompany the fast paced evolution of the inner layers. In contrast, the low impact regime is able to easily accommodate such measurement, thus, if a $\Delta \Pi$ value during the descending phase were derived from observations, -- in the conditions evaluated here -- it would favour axionic models.  In terms of the axion-photon coupling constant itself, the additional constraint in this scenario is not limiting enough to narrow its domain further. Despite not achieving such a low $\Delta \Pi$, the reference case still gave rise to multiple models with a $\chi^2$ close to 1.

\par
\begin{figure}[ht!]
    \centering
    \includegraphics[width=8.5cm]{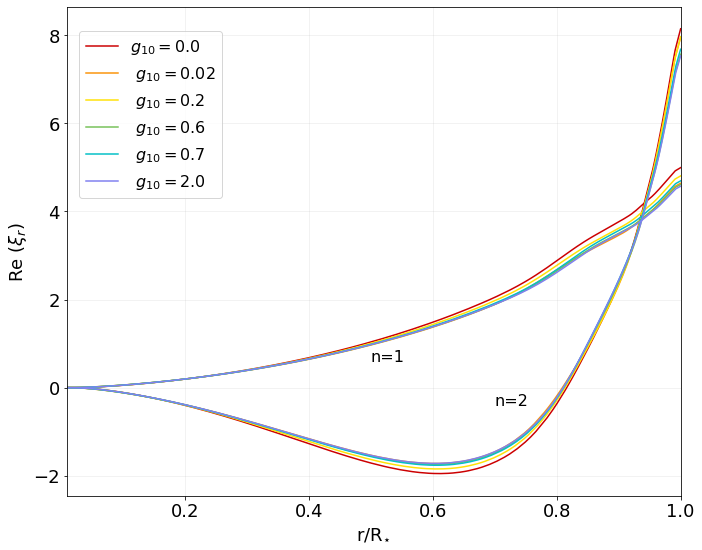}
    \caption{Radial displacement, $\xi_r$, in function of the radius, for the two observed modes. The curves identified with $n=1$ correspond to the fundamental mode and the ones labelled with $n=2$ represent the first overtone. }
    \label{fig:radial_modes}
\end{figure}

 Finally, we look at the radial displacement, $\xi_r$, that was calculated by GYRE \citep{GYRE2013} for the two observed radial modes. In Fig. \ref{fig:radial_modes}, the fundamental mode and first overtone are represented for the same selected models as in Figs. \ref{fig:3alpha} and \ref{fig:Enuc}. As expected, the two modes propagate throughout the whole star, however, they are most sensitive to the outer convective envelope, given that most of the red supergiant's mass is concentrated in the core -- $70\%$ of the mass is within a sphere with $25\%$ of its radius. Bearing in mind that the major changes in stellar structure during this phase of evolution as well as the impact of the axion energy losses are mainly concentrated in the most inner regions of the star, it is forthcoming that the behaviour of the two modes is similar for the models evaluated. Nevertheless, a slight decrease of the radial displacement occurs for the axion models in two occasions. For the first overtone, this is observed between $r = 0.4R_{\star}$ and $  r =0.8 R_{\star}$ ($m \in [0.79 ; \,0.96] \, M_{\star}$) while for the fundamental mode, this occurs mostly beyond $r=0.6R_{\star}$ ($m \geq 0.89\, M_{\star}$). That said, despite the penetrating power of these two modes of oscillation, they are not the best at probing the differences between the models evaluated here. The equivalent profile of mixed-modes would possibly be more revealing, since they transport information from the deeper layers, where g-modes are particularly sensitive, to the surface. Moving forward, we hope to perform a similar analysis for such modes that might arise in Betelgeuse.

\section{Conclusion}
\label{sec:conclusion}

\par As stated before, massive stars, and supergiants in particular, are still poorly explored territory, thus it is an ambitious task to set well established constraints on the axion-photon coupling constant based on their stellar behaviour. Nonetheless, in this work, we were able to produce the most reliable calibration yet for the reference stellar model as well for stellar axion models, using for the first time a combination of two mode frequencies, surface abundances, as well as the more typical quantities ($T_{eff}, \, \log L$ and $R$) as observational data. Even so, we obtained numerous possible models within a small interval of $\chi^2$, most with minimal differences from the input parameters presented in Tab. \ref{tab:modeloutputs}. It is fundamental to mention that further studies of Betelgeuse are likely to increase dramatically the confidence of our results, considering that estimates on characteristics such as age, mass and the big separation in period (related to the Brünt-Väisälä frequency profile) are able to exclude some of the models evaluated here. Moreover, additional observations of $\alpha$-Ori are likely to yield more precise measurements of quantities such as the radius, providing ground for improved versions of our models. Hopefully, this will be possible in the near future with missions such as PLATO. Nevertheless, even with the current available data, our results are in good agreement with other modelizations of Betelgeuse \citep[see e.g.][]{Luo2022, Dolan2016}. 
Thus, based on all the obtained models, with $g_{a\gamma} < 3.0 \times 10^{-10} {\rm GeV}^{-1}$, and $\chi^2 \leq 1.5$ (3912 samples), we favour the initial masses between $15M_{\odot}$ and  $18M_{\odot}$; 
initial metallicity between $-0.009< {\rm [FeH]_i} < -0.003$; initial He abundance between $0.21 < Y_i < 0.28$; mixing length parameter between $ 1.7< \alpha <  2.2$; and overshooting parameter between $0.015 < f_{ov} < 0.04$, given that none of our models converged to values outside these constraints. 
\par From another standpoint, this study contributes to the general knowledge of massive stars as well as the possible effects of axion energy losses in red supergiants. 
One argument in favour of this statement is the systematic increase in luminosity due to the increased rate of the 3$\alpha$ reactions along with the increased neutrino production. 
These are effects enhanced by the presence of axions, that are expected to be seen in similar stars. An increase of the overshooting parameter was also observed as the $g_{a \gamma}$ parameter became more meaningful, showing not only the effects of overshooting, but how different regimes favour different convection conditions. In regards to an upper bound on the $g_{a \gamma}$ constant, we set $g_{a \gamma } < 3.5 \times 10^{-10}{\rm GeV}^{-1}$, given that from this point on, the downhill simplex algorithm did not find any model that matched the observational data. A more restrictive upper limit would be $g_{a \gamma} < 3.0 \times 10^{-10}{\rm GeV}^{-1}$, based on the 
overshooting parameter above $0.055$, which is far beyond the typical accepted value, usually around $f_{ov}^{max} \approx 0.031$ \citep{Schootlemijer2019, Claret2017}.The latter argument can be particularly relevant since higher overshooting schemes are known to inhibit BSGs. Nevertheless, we adopt the more conservative upper limit, but keep in mind that most models achieved with $g_{10} = 3.0$ are not physically realistic. 
 Comparing our upper bound with other studies without strict mass limitations, such as \citet{Friedland2013}, our limit is less strict, however it compensates by its use a well calibrated red supergiant, instead of a general model, making it less contingent to stellar convection.
\par Finally, we should note that this work did not account for the effects of magnetic fields nor rotation -- which has been shown to affect surface abundances \citep{Luo2022}. The likeliness of the scenario in which Betelgeuse is the result of a past merger also limits our conclusions regarding this star, however, these considerations should not demean the analysis carried out here, as it provides useful insights about axions and massive stars. 

\begin{acknowledgments}
\par We thank Bill Paxton and all MESA contributors for making their code and work publicly available, including Alexander Friedland and Maurizio Giannotti for their contribution with the axion cooling patch. IL and CS thank the Funda\c c\~ao para a Ci\^encia e Tecnologia (FCT), Portugal, for the financial support to the Center for Astrophysics and Gravitation (CENTRA/IST/ULisboa)  through the Grant Project~No.~UIDB/00099/2020  and Grant No. PTDC/FIS-AST/28920/2017. 
\end{acknowledgments}
\newpage


\begin{thebibliography}{}
\providecommand{\natexlab}[1]{#1}
\providecommand{\url}[1]{\texttt{#1}}
\expandafter\ifx\csname urlstyle\endcsname\relax
\providecommand{\doi}[1]{doi: #1}\else
\providecommand{\doi}{doi: \begingroup \urlstyle{rm}\Url}\fi
	

\bibitem[Friedland et al.(2013)]{Friedland2013} Friedland, A., Giannotti, M., Wise, M. 2013, {\color{PineGreen}"Constraining the Axion-Photon Coupling with Massive Stars"}, \href{https://journals.aps.org/prl/abstract/10.1103/PhysRevLett.110.061101}{\color{gray}Physical Review Letters 110.}

\bibitem[Joyce et al.(2020)]{Joyce2020} Joyce, M., Leung, S.-C., Moln{\'a}r, L., Ireland, M., Kobayashi, C., Nomoto, K., 2020., {\color{PineGreen} "Standing on the Shoulders of Giants: New Mass and Distance Estimates for Betelgeuse through Combined Evolutionary, Asteroseismic, and Hydrodynamic Simulations with MESA"}, \href{https://iopscience.iop.org/article/10.3847/1538-4357/abb8db}{{\color{gray} The Astrophysical Journal 902.}}

\bibitem[Zwicky(1933)]{Zwicky1933} Zwicky, F., 1933., {\color{PineGreen}"Die Rotverschiebung von extragalaktischen Nebeln"}, {\color{gray}Helvetica Physica Acta 6, 110–127.}

\bibitem[Graham et al.(2015)]{Graham2015} Graham, P. W., Irastorza, I. G., Lamoreaux, S. K., Lindner, A., van Bibber, K. A., 2015., {\color{PineGreen} "Experimental Searches for the Axion and Axion-Like Particles"}, \href{https://www.annualreviews.org/doi/abs/10.1146/annurev-nucl-102014-022120}{{\color{gray}Annual Review of Nuclear and Particle Science 65, 485–514.}}


\bibitem[Peccei and Quinn(1977)]{PecceiQuinn1977} Peccei, R.~D., Quinn, H.~R., 1977., {\color{PineGreen} "CP conservation in the presence of pseudoparticles"}, \href{https://journals.aps.org/prl/abstract/10.1103/PhysRevLett.38.1440}{{\color{gray} Physical Review Letters 38, 1440–1443.}}

\bibitem[Raffelt(1990)]{Raffelt1990} Raffelt, G.~G., 1990., {\color{PineGreen} "Astrophysical methods to constrain axions and other novel particle phenomena"}, \href{https://www.sciencedirect.com/science/article/abs/pii/0370157390900546}{{\color{gray}Physics Reports 198, 1–113.}} 


\bibitem[Vinyoles et al.(2015)]{Vinyoles2015} Vinyoles, N., Serenelli, A., Villante, F.~L., Basu, S., Redondo, J., Isern, J., 2015, {\color{PineGreen}"New axion and hidden photon constraints from a solar data global fit"}, \href{https://iopscience.iop.org/article/10.1088/1475-7516/2015/10/015}{{\color{gray} Journal of Cosmology and Astroparticle Physics 2015.}} 

\bibitem[Anastassopoulos et al.(2017)]{CAST(2017)} Anastassopoulos, V. and 65 colleagues 2017.{\color{PineGreen} "New CAST limit on the axion-photon interaction"}, \href{https://www.nature.com/articles/nphys4109}{{\color{gray}Nature Physics 13, 584–590.}}

\bibitem[Armengaud et al.(2019)]{IAXO(2019)} Armengaud, E. and 90 colleagues, 2019, {\color{PineGreen}"Physics potential of the International Axion Observatory (IAXO)"}, \href{https://iopscience.iop.org/article/10.1088/1475-7516/2015/02/006}{{\color{gray}Journal of Cosmology and Astroparticle Physics 2019.}} 

\bibitem[Ringwald et al.(2020)]{pdg2020} . Ringwald, L.J. Rosenberg and G. Rybka, 2020, {\color{PineGreen}"Axions and Other Similar Particles"}, \href{P.A. Zyla et al. (Particle Data Group), Prog. Theor. Exp. Phys. 2020, 083C01 (2020)}{{\color{gray}Particle Data Group}} 


\bibitem[Xiao et al.(2021)]{Xiao2021} Xiao, M. and 7 colleagues 2021, {\color{PineGreen} "Constraints on Axionlike Particles from a Hard X-Ray Observation of Betelgeuse"}, \href{https://journals.aps.org/prl/abstract/10.1103/PhysRevLett.126.031101}{{\color{gray} Physical Review Letters 126.}} 


\bibitem[Haubois et al.(2009)]{Haubois2009} Haubois, X. and 11 colleagues 2009, {\color{PineGreen}"Imaging the spotty surface of Betelgeuse in the H band"}, \href{https://www.aanda.org/articles/aa/pdf/2009/47/aa12927-09.pdf}{{\color{gray}Astronomy and Astrophysics 508, 923–932.}} 


\bibitem[Kiss et al.(2006)]{Kiss2006} Kiss, L.~L., Szab{\'o}, G.~M., Bedding, T.~R., 2006., {\color{PineGreen}"Variability in red supergiant stars: pulsations, long secondary periods and convection noise"}, \href{https://academic.oup.com/mnras/article/372/4/1721/1189711}{{\color{gray} Monthly Notices of the Royal Astronomical Society 372, 1721–1734.}} 

\bibitem[Lambert et al.(1984)]{Lambert1984} Lambert, D.~L., Brown, J.~A., Hinkle, K.~H., Johnson, H.~R., 1984., {\color{PineGreen} " Carbon, nitrogen and oxygem abundances in Betelgeuse"}, \href{https://ui.adsabs.harvard.edu/abs/1984ApJ...284..223L/abstract}{{\color{gray} The Astrophysical Journal 284, 223–237.}} 



\bibitem[Dolan et al.(2016)]{Dolan2016} Dolan, M.~M., Mathews, G.~J., Lam, D.~D., Quynh Lan, N., Herczeg, G.~J., Dearborn, D.~S.~P., 2016., {\color{PineGreen}"Evolutionary Tracks for Betelgeuse"}, \href{https://iopscience.iop.org/article/10.3847/0004-637X/819/1/7}{{\color{gray}The Astrophysical Journal 819.}} 

\bibitem[Nance et al.(2018)]{Nance2018} Nance, S., Sullivan, J.~M., Diaz, M., Wheeler, J.~C, 2018, {\color{PineGreen} "The Betelgeuse Project II: asteroseismology"}, \href{https://academic.oup.com/mnras/article/479/1/251/5032801}{{\color{gray}Monthly Notices of the Royal Astronomical Society 479, 251–261.}} 

\bibitem[Neilson et al.(2011)]{Neilson2011} Neilson, H.~R., Lester, J.~B., Haubois, X., 2011, {\color{PineGreen}"Weighing Betelgeuse: Measuring the Mass of {\ensuremath{\alpha}} Orionis from Stellar Limb-darkening"}, \href{https://arxiv.org/abs/1109.4562}{{\color{gray}9th Pacific Rim Conference on Stellar Astrophysics 451, 117.}}

\bibitem[Wheeler et al.(2017)]{Wheeler2017} Wheeler, J.~C. and 8 colleagues 2017, {\color{PineGreen} "The Betelgeuse Project: constraints from rotation"}, \href{https://academic.oup.com/mnras/article/465/3/2654/2454753}{{\color{gray}Monthly Notices of the Royal Astronomical Society 465, 2654–2661.}} 

\bibitem[Townsend and Teitler(2013)]{GYRE2013} Townsend, R.~H.~D., Teitler, S.~A., 2013, {\color{PineGreen} "GYRE: an open-source stellar oscillation code based on a new Magnus Multiple Shooting scheme"}, \href{https://academic.oup.com/mnras/article/435/4/3406/1033475}{{\color{gray}Monthly Notices of the Royal Astronomical Society 435, 3406–3418.}} 

\bibitem[Bertone and Hooper(2018)]{Bertone2018} Bertone, G., Hooper, D., 2018., {\color{PineGreen}"History of dark matter"}, \href{https://journals.aps.org/rmp/abstract/10.1103/RevModPhys.90.045002}{{\color{gray} Reviews of Modern Physics 90.}}

\bibitem[Jungman et al.(1996)]{Jungman1996} Jungman, G., Kamionkowski, M., Griest, K., 1996, {\color{PineGreen}"Supersymmetric dark matter"} \href{https://www.sciencedirect.com/science/article/abs/pii/0370157395000585?via\%3Dihub}{{\color{gray}Physics Reports 267, 195–373.}} 

\bibitem[Dasgupta and Kopp(2021)]{Dasgupta2021} Dasgupta, B., Kopp, J., 2021, {\color{PineGreen}"Sterile neutrinos"} \href{https://www.sciencedirect.com/science/article/pii/S0370157321002696?via\%3Dihub}{{\color{gray}Physics Reports 928, 1–63.}} 

\bibitem[Abel et al.(2020)]{Abel2020} Abel, C. and 83 colleagues, 2020, {\color{PineGreen}"Measurement of the Permanent Electric Dipole Moment of the Neutron"} \href{https://journals.aps.org/prl/abstract/10.1103/PhysRevLett.124.081803}{{\color{gray}Physical Review Letters 124.}} 

\bibitem[Weinberg(1978)]{Weinberg1978} Weinberg, S., 1978., {\color{PineGreen}"A new light boson?"} \href{https://journals.aps.org/prl/abstract/10.1103/PhysRevLett.40.223}{{\color{gray}Physical Review Letters 40, 223–226.}} 

\bibitem[Marsh(2016)]{Marsh2016} Marsh, D.~J.~E., 2016, {\color{PineGreen}"Axion cosmology"} \href{https://www.sciencedirect.com/science/article/abs/pii/S0370157316301557}{{\color{gray}Physics Reports 643, 1–79.}} 


\bibitem[Choplin et al.(2017)]{Choplin2017} Choplin, A., Coc, A., Meynet, G., Olive, K.~A., Uzan, J.-P., Vangioni, E.\ 2017, {\color{PineGreen}"Effects of axions on Population III stars"}, \href{https://www.aanda.org/articles/aa/full_html/2017/09/aa31040-17/aa31040-17.html}{{\color{gray}Astronomy and Astrophysics 605.}} 


\bibitem[Raffelt(2008)]{Raffelt2008} Raffelt, G.~G.\ 2008, {\color{PineGreen}"Astrophysical Axion Bounds"}, \href{http://www.icepp.s.u-tokyo.ac.jp/~asai/work/LN_Axion.pdf}{{\color{gray}Axions 51.}}


\bibitem[Levesque and Massey(2020)]{Levesque2020} Levesque, E.~M., Massey, P.\ 2020, {\color{PineGreen}"Betelgeuse Just Is Not That Cool: Effective Temperature Alone Cannot Explain the Recent Dimming of Betelgeuse"}, \href{https://iopscience.iop.org/article/10.3847/2041-8213/ab7935/meta}{{\color{gray}The Astrophysical Journal 891.}} 

\bibitem[Paxton et al.(2011)]{Paxton2011} Paxton, B., Bildsten, L., Dotter, A., Herwig, F., Lesaffre, P., Timmes, F.\ 2011, {\color{PineGreen}"Modules for Experiments in Stellar Astrophysics (MESA)"}, \href{https://iopscience.iop.org/article/10.1088/0067-0049/192/1/3}{{\color{gray}The Astrophysical Journal Supplement Series 192.}} 

\bibitem[Paxton et al.(2013)]{Paxton2013} Paxton, B. and 10 colleagues 2013, {\color{PineGreen}"Modules for Experiments in Stellar Astrophysics (MESA): Planets, Oscillations, Rotation, and Massive Stars"}, \href{https://iopscience.iop.org/article/10.1088/0067-0049/208/1/4}{{\color{gray}The Astrophysical Journal Supplement Series 208.}} 

\bibitem[Paxton et al.(2015)]{Paxton2015} Paxton, B. and 12 colleagues 2015, {\color{PineGreen}"Modules for Experiments in Stellar Astrophysics (MESA): Binaries, Pulsations, and Explosions"}, \href{https://iopscience.iop.org/article/10.1088/0067-0049/220/1/15}{{\color{gray}The Astrophysical Journal Supplement Series 220.}} 

\bibitem[Paxton et al.(2018)]{Paxton2018} Paxton, B. and 12 colleagues 2018, {\color{PineGreen}"Modules for Experiments in Stellar Astrophysics (MESA): Convective Boundaries, Element Diffusion, and Massive Star Explosions"}, \href{https://iopscience.iop.org/article/10.3847/1538-4365/aaa5a8}{{\color{gray}The Astrophysical Journal Supplement Series 234.}} 


\bibitem[Paxton et al.(2019)]{Paxton2019} Paxton, B. and 16 colleagues 2019, {\color{PineGreen}"Modules for Experiments in Stellar Astrophysics (MESA): Pulsating Variable Stars, Rotation, Convective Boundaries, and Energy Conservation"}, \href{https://iopscience.iop.org/article/10.3847/1538-4365/ab2241}{{\color{gray}The Astrophysical Journal Supplement Series 243.}} 


\bibitem[Nelder and Mead (1965)]{Nelder1965} Nelder A. J. and Mead R.\ 1965, {\color{PineGreen}"A Simplex Method for Function Minimization"}, \href{https://academic.oup.com/comjnl/article-abstract/7/4/308/354237}{{\color{gray}The Computer Journal, Volume 7, Issue 4.}}

\bibitem[Kippenhahn et al.(2012)]{Kippenhahn2012} Kippenhahn, R., Weigert, A., Weiss, A.\ 2012.\ {\color{PineGreen}Stellar Structure and Evolution.}\ \href{https://link.springer.com/book/10.1007/978-3-642-30304-3}{{\color{gray}Stellar Structure and Evolution, ISBN: 978-3-642-30304-3. Berlin, Heidelberg: Springer Berlin Heidelberg, 2012.}} 

\bibitem[Aerts et al.(2010)]{Aerts2010} Aerts, C., Christensen-Dalsgaard, J., Kurtz, D.~W.\ 2010.\ {\color{PineGreen}Asteroseismology.}\ \href{https://link.springer.com/book/10.1007/978-1-4020-5803-5}{{\color{gray} Asteroseismology, Astronomy and Astrophysics Library. ISBN 978-1-4020-5178-4. Springer Science+Business Media B.V., 2010, p..}}

\bibitem[Raffelt(1999)]{Raffelt1999} Raffelt, G.\ 1999.\ {\color{PineGreen}"Stellar-evolution limits on axion properties"}.\ \href{https://www.sciencedirect.com/science/article/abs/pii/S0920563298005015?via\%3Dihub}{{\color{gray}Nuclear Physics B Proceedings Supplements 72, 43–53.}} 

\bibitem[Schootemeijer et al.(2019)]{Schootlemijer2019} Schootemeijer, A., Langer, N., Grin, N.~J., Wang, C.\ 2019.\ {\color{PineGreen}"Constraining mixing in massive stars in the Small Magellanic Cloud"}, \href{https://www.aanda.org/articles/aa/full_html/2019/05/aa35046-19/aa35046-19.html}{\color{gray}Astronomy and Astrophysics 625.} 

\bibitem[Bowman(2020)]{Bowman2020} Bowman, D.~M.\ 2020, {\color{PineGreen}"Asteroseismology of high-mass stars: new insights of stellar interiors with space telescopes"} \href{https://www.frontiersin.org/articles/10.3389/fspas.2020.578584/full}{{\color{gray}Frontiers in Astronomy and Space Sciences 7.}} 

\bibitem[Rauer et al.(2021)]{Rauer2021} Rauer, H. and 12 colleagues 2021.\ {\color{PineGreen}"The PLATO mission: Overview and status"}, \href{https://meetingorganizer.copernicus.org/EPSC2021/EPSC2021-90.html}{{\color{gray}European Planetary Science Congress.}}

\bibitem[Cantiello et al.(2021)]{Cantiello2021} Cantiello, M., Lecoanet, D., Jermyn, A.~S., Grassitelli, L.\ 2021.\ {\color{PineGreen}"On the Origin of Stochastic, Low-Frequency Photometric Variability in Massive Stars"} \href{https://www.scholars.northwestern.edu/en/publications/on-the-origin-of-stochastic-low-frequency-photometric-variability}{{\color{gray}The Astrophysical Journal 915.}} 


\bibitem[Di Mauro(2016)]{Mauro2016} Di Mauro, M.~P.\ 2016.\ {\color{PineGreen}"A review on Asteroseismology"} \href{https://www.semanticscholar.org/paper/A-review-on-Asteroseismology-Mauro/55ccd4fdaea100a238f84c205eb203b7c5ae49c6}{{\color{gray}Frontier Research in Astrophysics II (FRAPWS2016).}}

\bibitem[Montalb{\'a}n et al.(2013)]{Montalban2013} Montalb{\'a}n, J., Miglio, A., Noels, A., Dupret, M.-A., Scuflaire, R., Ventura, P., 2013, {\color{PineGreen}"Testing Convective-core Overshooting Using Period Spacings of Dipole Modes in Red Giants"}, \href{https://research.birmingham.ac.uk/en/publications/testing-convective-core-overshooting-using-period-spacings-of-dip}{{\color{gray}The Astrophysical Journal 766.}} 

\bibitem[Gray(2008)]{Gray2008} Gray, D.~F., 2008, {\color{PineGreen}"Mass Motions in the Photosphere of Betelgeuse"}, \href{https://iopscience.iop.org/article/10.1088/0004-6256/135/4/1450}{{\color{gray}The Astronomical Journal 135, 1450–1458.}} 

\bibitem[Chatzopoulos et al.(2020)]{Chatzopoulos2020} Chatzopoulos, E. Frank, J., Marcello, D.~C., 2020, {\color{PineGreen}"Is Betelgeuse the Outcome of a Past Merger?"}, \href{https://iopscience.iop.org/article/10.3847/1538-4357/ab91bb}{{\color{gray}The Astrophysical Journal 896.}} 

\bibitem[Sullivan et al.(2020)]{Sullivan2020} Sullivan, J.~M., Nance, S., Wheeler, J.~C., 2020, {\color{PineGreen}"The Betelgeuse Project. III. Merger Characteristics"}, \href{https://iopscience.iop.org/article/10.3847/1538-4357/abc3c9}{{\color{gray}The Astrophysical Journal 905.}} 

\bibitem[Meynet et al.(2013)]{Meynet2013} Meynet, G., Haemmerl{\'e}, L., Ekstr{\"o}m, S., Georgy, C., Groh, J., Maeder, A, 2013, {\color{PineGreen}"The past and future evolution of a star like Betelgeuse"}, \href{https://www.eas-journal.org/articles/eas/abs/2013/02/eas1360002/eas1360002.html}{{\color{gray}EAS Publications Series 60, 17–28.}} 

\bibitem[Kravchenko et al.(2021)]{Kravchenko2021} Kravchenko, K. and 9 colleagues 2021, {\color{PineGreen}"Atmosphere of Betelgeuse before and during the Great Dimming event revealed by tomography"}, \href{https://www.aanda.org/articles/aa/pdf/2021/06/aa39801-20.pdf}{{\color{gray}Astronomy and Astrophysics 650.}} 

\bibitem[Luo et al.(2022)]{Luo2022} Luo, T., Umeda, H., Yoshida, T., Takahashi, K., 2022, {\color{PineGreen}"Stellar Models of Betelgeuse Constrained Using Observed Surface Conditions"}, \href{https://iopscience.iop.org/issue/0004-637X/927/1}{{\color{gray}The Astrophysical Journal 927.}} 


\bibitem[Duffy and van Bibber(2009)]{Duffy2009} Duffy, L.~D., van Bibber, K., 2009, {\color{PineGreen}"Axions as dark matter particles"} \href{https://iopscience.iop.org/article/10.1088/1367-2630/11/10/105008}{{\color{gray}New Journal of Physics 11.}} 

\bibitem[Yang and di(2017)]{Yang2017} Yang, Q., di, H., 2017, {\color{PineGreen}"Axion-like particle dark matter in the linear regime of structure formation"}, \href{https://www.worldscientific.com/doi/abs/10.1142/S0217751X17500518}{{\color{gray}International Journal of Modern Physics A 32.}} 

\bibitem[Agrawal et al.(2020)]{Agrawal2020} Agrawal, P., Hurley, J., Stevenson, S., Sz{\'e}csi, D., Flynn, C., 2020, {\color{PineGreen}"The fates of massive stars: exploring uncertainties in stellar evolution with METISSE"}, \href{https://academic.oup.com/mnras/article-abstract/497/4/4549/5881964?redirectedFrom=PDF}{{\color{gray}}Monthly Notices of the Royal Astronomical Society 497, 4549–4564.} 

\bibitem[Claret and Torres(2017)]{Claret2017} Claret, A., Torres, G, 2017, {\color{PineGreen}"The Dependence of Convective Core Overshooting on Stellar Mass: A Semi-empirical Determination Using the Diffusive Approach with Two Different Element Mixtures"}, \href{https://iopscience.iop.org/article/10.3847/1538-4357/aa8770}{{\color{gray}The Astrophysical Journal 849.}}

\bibitem[Pedersen et al.(2018)]{Pedersen2018} Pedersen, M.~G., Aerts, C., P{\'a}pics, P.~I., Rogers, T.~M., 2018, {\color{PineGreen}"The shape of convective core overshooting from gravity-mode period spacings"}, \href{https://www.aanda.org/articles/aa/full_html/2018/06/aa32317-17/aa32317-17.html}{{\color{gray}Astronomy and Astrophysics 614.}}

\bibitem[Metcalfe et al.(2012)]{Metcalfe2012} Metcalfe, T.~S. and 36 colleagues 2012, {\color{PineGreen}"Asteroseismology of the Solar Analogs 16 Cyg A and B from Kepler Observations"}, \href{https://iopscience.iop.org/article/10.1088/2041-8205/748/1/L10}{{\color{gray}The Astrophysical Journal 748.}} 

\bibitem[Cowling(1941)]{Cowling1941} Cowling, T.~G., 1941, {\color{PineGreen}"The non-radial oscillations of polytropic stars"}, \href{https://academic.oup.com/mnras/article/101/8/367/2600915}{{\color{gray}Monthly Notices of the Royal Astronomical Society 101, 367.}} 

\bibitem[Unno et al.(1979)]{Unno1979} Unno, W., Osaki, Y., Ando, H., Shibahashi, H., 1979, {\color{PineGreen}"Nonradial oscillations of stars"}, \href{}{{\color{gray}Tokyo: University of Tokyo Press, 1979.}}

\bibitem[Delholm et al.(2012)]{Delholm2012} Dehlholm, C. and Brockhoff, P. B. and  Bredie, W. L. P, 2012 {\color{PineGreen} "Confidence ellipses: A variation based on parametric bootstrapping applicable on Multiple Factor Analysis results for rapid graphical evaluation"}, \href{https://www.sciencedirect.com/science/article/pii/S0950329312000766}{{\color{gray} Food Quality and Preference 26, 2.}}

\bibitem[Planck Collaboration et al.(2016)]{Planck2016} Planck Collaboration and 261 colleagues 2016.\ {\color{PineGreen}Planck 2015 results. XIII. Cosmological parameters.}\ \href{https://www.aanda.org/articles/aa/full_html/2016/10/aa25830-15/aa25830-15.html}{{\color{gray} Astronomy and Astrophysics 594. }}

\end{thebibliography}

\bibliographystyle{aasjournal}



\end{document}